# On Clustering Trend in Language Evolution Based on Dynamical Behaviors of Multi-Agent Model

Yu Zhang[1], Li Liu[1], Chen Diao[2], Ning Cai*[1]

[1] School of Artificial Intelligence, Beijing University of Posts and Telecommunications, Beijing, China
E-mail: caining91@tsinghua.org.cn

[2] School of Information Engineering, Ningxia University, Yinchuan, China

**Abstract:** Computer model has been extensively adopted to overcome the time limitation of language evolution by transforming language theory into physical modeling mechanism, which helps to explore the general laws of the evolution. In this paper, a multi-agent model is designed to simulate the evolution process of language in human settlements, with the network topology being lattice. The language of each node in the lattice will evolve gradually under the influence of its own fixed evolutionary direction and neighbors. According to the computational experiment results, it is discovered that the state points of languages always converge into several clusters during evolution process, which gives us an insight into language evolution.

## I. INTRODUCTION

Human language is unique among communication systems in nature: it is socially learned and offers the potential for open communication because of its recursive constituent structure [1]. The structure of this communication system is not only the result of biological ability evolution but also that of cultural evolution of language itself [2]. In general, the research purpose of language evolution is: when, where, and how human language came into being, changed, and died [3], with the objects including the meaning of language, grammar, and other details [4-6].

Due to the lack of sufficient data support and the extremely slow speed of language evolution in reality, the most commonly appliable method for historical research on language evolution is based on computational model [7-9]. One research paradigm is inspired by the idea that language may arise spontaneously in a communicative group and may hold some adaptive features [10]. This view has led to a great deal of analysis of multi-agent models that simulate such communication and seek to infer the attributes of emerging languages and their possible further evolution [11-13].

According to the mathematical realization of the components of the target language phenomenon, existing language models can be classified to rule-based and equation-based models [14-15]. Rule-

---

based models define concrete or abstract rules to describe or operate language components and related behaviors, with the interrelation of these rules leading to evolution [16-17]. Equation-based models tend to convert languages and related behaviors into mathematical equations, and mathematical analysis and experimental or empirical verification of these equations enable equation-based models to reasonably approximate the history of language evolution or forecast its future [18]. Many computational models are proposed based on the above two frameworks. For example, some models attribute the distribution of phonetic elements to self-organization in the communication process [19-20], that is, the process in which the global pattern of the system is generated from the local interaction of its components [21]. Some studies have also proved that the universality of language is naturally generated through cultural transmission [22-24] or language games [25-28].

In this paper, we design a multi-agent model to simulate the evolution of language over time. First, the basic unit of language transmission and exchange should be a human settlement and it will act as an agent [29]. To construct the framework of language communication and variance, we take each human settlement as a node in the network topology and construct the configuration of topology between nodes inspired by ecological competition models. Each node in the network has two basic attributes: weight and state, which correspond to the degree of influence of settlements and the development state of language and culture respectively. In addition, connecting edge of any two nodes has a basic attribute of edge weight, which represents the intensity of traffic flow between any two settlements. As a basic rule of thumb, for example, the higher the weight of a node, the greater the force it exerts on the surrounding nodes with which it is connected. Then, a language dynamics model is designed to simulate the linguistic and cultural development of settlements by integrating relevant principles of multi-disciplines [30], that is, the states of all nodes in the topological network are iteratively changed through the dynamics model.

Based on the above model, we conducted computer simulation [31] experiments to study the influence of configuration of topology, variation of state dimensions, and other factors on language evolution and found an adaptive feature in language evolution. We conclude that this model reasonably models the dynamics of language evolution by analyzing the feedback of experimental results and combining relevant empirical conclusions. Moreover, it reveals the inherent commonality between linguistic, physical, and biological phenomena, since the model itself and its mathematical principles can reflect a more extensive and general background other than the domain of linguistics. In brief, this innovative language evolution model enriches our mentality and technique of studying language evolution.

The rest of this paper is organized as follows. Sec. 2 elaborates and formulates the model and proves the condition of relative stability theoretically. Four computational experiments are carried out to explore the general laws of language evolution in Sec. 3. Finally, Sec. 4 draws the conclusion and indicates future directions.

**II. MODEL CONSTRUCTION**

The unit of language evolution is residential settlement with fixed position, possibly being a city, town, or a village. In ancient society, the communication between any two neighboring settlements could only rely on the road between them because of the restriction of single form of communication. Furthermore, the communication between non-adjacent settlements needs to go through multiple transfer stations, without direct communication. Therefore, the network topology between settlements must be a planar graph. For simplicity, it is assumed to be a regular lattice, as shown in Fig. 1:

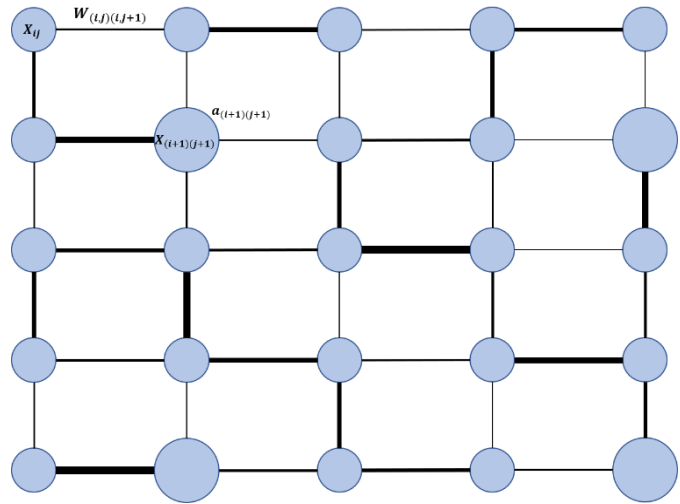

FIGURE 1. Network topology diagram of settlements.

Nodes in the network topology represent settlements and edges represent roads between them. Since the nodes are in the lattice, the number of each node is denoted by a dyadic array $(i,j)$. For any node in the network, its language evolution mechanism can be summarized as the following formula:

$$x_{ij}(k+1) = \underbrace{x_{ij}(k) + d_{ij}}_{Self-Evolution} + \underbrace{\sum_{(m,n)\in N_{ij}} w_{(i,j)(m,n)} a_{mn} A \Delta_{(i,j)(m,n)}(k)}_{Interactive\ Influence} \qquad (1)$$

In this formula, $k$ denotes time, and $x_{ij}(k) \in R^p$ denotes the $p$-dimensional state vector of the settlement $(i,j)$. Each entry of the state vector quantitatively corresponds to certain feature of the language, and an instantaneous portrait of the evolution of language can be expressed by a dynamic high-dimensional state vector stacked by a set of such entries. $d_{ij} \in R^p$ is the self-evolution offset, which is generated randomly initially. $A$ denotes the coupling matrix, which is used to describe the dynamically coupling relationship between the entries of the state vector. $N_{ij}$ is the neighborhood of node $(i,j)$, which contains all paths directly connected to the point $(i,j)$. $a_{ij} \in R^+$ indicates the degree of influence of the node $(i,j)$. $\Delta_{(i,j)(m,n)}(k) = x_{mn}(k) - x_{ij}(k)$. $w_{(i,j)(m,n)} \in R^+$ represents the edge weight of the network, and its connotation is the intensity of connectivity between nodes $(i,j)$ and $(m,n)$, which depends on the unimpeded degree of the actual road between the two settlements. Therefore, the network topology is an undirected graph, namely:

$$w_{(i,j)(m,n)} = w_{(m,n)(i,j)} \tag{2}$$

In particular, when the linguistic and cultural differences between any two neighboring settlements reach a certain threshold, they will also be unable to communicate, even if there is a physical pathway between them. In the model, if

$$\left\|\Delta_{(i,j)(m,n)}(k)\right\|_\infty > \psi \tag{3}$$

then $\Delta_{(i,j)(m,n)}(k)$ is forcibly adjusted to 0. and $\psi$ denotes the threshold.

The dynamics of any node consist of two parts: self-evolution and interaction between nodes. The impetus of interaction is proportional to the weight of nodes and the strength of connections between nodes, while the impetus of self-evolution is determined by the state vector and the offset of self-evolution. Each node in the network topology evolves according to the above evolution mechanism, after setting certain appropriate initial values of variables. In this way, our language evolution model is designed.

A basic premise for a language evolution model to work properly is stability, that is, to ensure that the states of settlements do not diverge from each other. Without loss of generality, take two specific nodes as an example, and their connection is shown in the following figure:

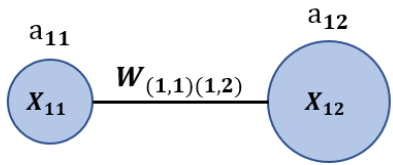

FIGURE 2. Graph topology of a two-settlement system.

In Fig. 2, the dynamic equation of node $(1,1)$ is:

$$x_{11}(k+1) = x_{11}(k) + d_{11} + w_{(1,1)(1,2)}a_{12}A\Delta_{(1,1)(1,2)}(k) \tag{4}$$

Similarly, the dynamic equation of node $(1,2)$ is:

$$x_{12}(k+1) = x_{12}(k) + d_{12} + w_{(1,1)(1,2)} a_{11} A \Delta_{(1,2)(1,1)}(k) \quad (5)$$

Let $\Delta_{(1,1)(1,2)}(k) = (x_{12}(k) - x_{11}(k))$, then the dynamic equation of $\Delta_{(1,1)(1,2)}(k)$ can be derived by subtracting (4) from (5) as:

$$\begin{aligned}\Delta_{(1,1)(1,2)}(k+1) &= \Delta_{(1,1)(1,2)}(k) + \Delta d + w_{(1,1)(1,2)} A(-a_{11} - a_{12})\Delta_{(1,1)(1,2)}(k) \\ &= \underbrace{[I - w_{(1,1)(1,2)}(a_{11} + a_{12})A]\Delta_{(1,1)(1,2)}(k)}_{Exponential\ growth} + \underbrace{\Delta d}_{Linear\ cumulation}\end{aligned} \quad (6)$$

For the above differential equation, the linear accumulation part can be regarded as disturbance and does not affect the stability, which is determined by the exponential change part. Therefore, the stability of the equation can be described by the following theorem:

**THEOREM 1**

For the dynamical system (6), if $\lambda_1, \lambda_2, \ldots, \lambda_n$ are the eigenvalues of

$$I - w_{(1,1)(1,2)}(a_{11} + a_{12})A \quad (7)$$

Then a necessary and sufficient condition for the asymptotic stability is

$$\rho(I - w_{(1,1)(1,2)}(a_{11} + a_{12})A) = \max_{1 \le i \le n}|\lambda_i| < 1 \quad (8)$$

where $\rho(\cdot)$ is the spectral radius. When $A = I$, the condition can be simplified as:

$$w_{(1,1)(1,2)}(a_{11} + a_{12}) < 1 \quad (9)$$

**PROOF:**

Sufficiency. Let $\rho(I - w_{(1,1)(1,2)}(a_{11} + a_{12})A) < 1$, then we have

$$|1 - \lambda(w_{(1,1)(1,2)}(a_{11} + a_{12})A)| < 1 \quad (10)$$

where $\lambda(\cdot)$ denotes any eigenvalue of the matrix.

The inequality means $\lambda(w_{(1,1)(1,2)}(a_{11} + a_{12})A) \ne 0$, namely the matrix $w_{(1,1)(1,2)}(a_{11} + a_{12})A$ is nonsingular. Therefore, one can derive the conclusion that the equation

$$w_{(1,1)(1,2)}(a_{11} + a_{12})A\Delta_{(1,1)(1,2)} = \Delta d \quad (11)$$

has a unique solution. The solution can be written as $\Delta_{(1,1)(1,2)}^*$

$$\Delta_{(1,1)(1,2)}^* = [I - w_{(1,1)(1,2)}(a_{11} + a_{12})A]\Delta_{(1,1)(1,2)}^* + \Delta d \quad (12)$$

Then the error vector is

$$\begin{aligned}
\varepsilon(k) &= \Delta_{(1,1)(1,2)}(k) - \Delta_{(1,1)(1,2)}{}^* = [I - w_{(1,1)(1,2)}(a_{11} + a_{12})A]\Delta_{(1,1)(1,2)}(k-1) \\
&\quad + \Delta d - \Delta_{(1,1)(1,2)}{}^* \\
&= [I - w_{(1,1)(1,2)}(a_{11} + a_{12})A](\Delta_{(1,1)(1,2)}(k-1) - \Delta_{(1,1)(1,2)}{}^*) \\
&= [I - w_{(1,1)(1,2)}(a_{11} + a_{12})A]^2(\Delta_{(1,1)(1,2)}(k-2) - \Delta_{(1,1)(1,2)}{}^*) \\
&\quad \cdots \\
&= [I - w_{(1,1)(1,2)}(a_{11} + a_{12})A]^k(\Delta_{(1,1)(1,2)}(0) - \Delta_{(1,1)(1,2)}{}^*) \\
&= [I - w_{(1,1)(1,2)}(a_{11} + a_{12})A]^k \varepsilon(0)
\end{aligned}$$

(13)

where $\varepsilon(0) = \Delta_{(1,1)(1,2)}(0) - \Delta_{(1,1)(1,2)}{}^*$.

We might as well assume that $I - w_{(1,1)(1,2)}(a_{11} + a_{12})A$ is diagonalizable [32], then we can set

$$P^{-1}[I - w_{(1,1)(1,2)}(a_{11} + a_{12})A]P = \begin{bmatrix} \lambda_1 & 0 & \cdots & 0 \\ 0 & \lambda_2 & \cdots & 0 \\ \vdots & \vdots & \ddots & \vdots \\ 0 & 0 & \cdots & \lambda_n \end{bmatrix} = \Lambda \tag{14}$$

From this, it naturally yields that

$$\begin{aligned}[I - w_{(1,1)(1,2)}(a_{11} + a_{12})A]^k &= P\Lambda P^{-1} P\Lambda P^{-1} \cdots P\Lambda P^{-1} = P\Lambda^k P^{-1} \\
&= P \begin{bmatrix} \lambda_1^k & 0 & \cdots & 0 \\ 0 & \lambda_2^k & \cdots & 0 \\ \vdots & \vdots & \ddots & \vdots \\ 0 & 0 & \cdots & \lambda_n^k \end{bmatrix} P^{-1}
\end{aligned}$$

(15)

The equation

$$\lim_{k \to \infty} \lambda_1^k = \lim_{k \to \infty} \lambda_2^k = \cdots = \lim_{k \to \infty} \lambda_n^k = 0 \tag{16}$$

is true because Eq. (8) holds. Therefore, we have

$$\lim_{k \to \infty}[I - w_{(1,1)(1,2)}(a_{11} + a_{12})A]^k = 0 \tag{17}$$

Then for any $\Delta_{(1,1)(1,2)}(0)$, we have $\lim_{k \to \infty} \varepsilon(k) = 0$, i.e.

$$\lim_{k \to \infty} \Delta_{(1,1)(1,2)}(k) = \Delta_{(1,1)(1,2)}{}^* \tag{18}$$

Necessity. If $[I - w_{(1,1)(1,2)}(a_{11} + a_{12})A]$ has an eigenvalue $\lambda^*$, then the condition

$$[I - w_{(1,1)(1,2)}(a_{11} + a_{12})A]x = \lambda^* x \tag{19}$$

is satisfied for certain $x$, where $|\lambda^*| \geq 1$ and $x \neq 0$. From this, one can obtain the conclusion that

$$\left\|[I - w_{(1,1)(1,2)}(a_{11} + a_{12})A]^k x\right\| = |\lambda^*|^k \|x\| \tag{20}$$

Therefore, $[I - w_{(1,1)(1,2)}(a_{11} + a_{12})A]^k x$ does not converge to the zero vector when $k \to \infty$. For the inequality

$$\left\| [I - w_{(1,1)(1,2)}(a_{11} + a_{12})A]^k x \right\| \leq \left\| [I - w_{(1,1)(1,2)}(a_{11} + a_{12})A]^k \right\| \|x\| \quad (21)$$

is true, $[I - w_{(1,1)(1,2)}(a_{11} + a_{12})A]^k$ does not converge to the matrix zero, and

$$\lim_{k \to \infty} \Delta_{(1,1)(1,2)}(k) \neq \Delta_{(1,1)(1,2)}^{*} \quad (22)$$

can be concluded from Eq. (13).□

Because the isolated two-settlement system is a subsystem of the whole dynamical network, the stability of the two-settlement system is a necessary condition for the stability of the system overall.

## III. EXPERIMENT AND ANALYSIS

In the language evolution model, some variables that can affect the results of language evolution include self-evolution offset, edge and node weight, and dimensions of the state vector. Through experiments, we will reveal the causal effects between these influencing factors and the results of language evolution, in order to enrich our knowledge of language evolution dynamics.

### A. SETTING OF PARAMETERS

Before the experiment, we need to set the relevant parameters first. The network topology is set as a 20×20 grid, with each node having a time-varying state. The state vector and self-evolution offset are both two-dimensional vectors. The initial value of each entry of the state vector is randomly generated within (0, 2), with that of the offset vector within (-0.02, 0.08). The weights of the edges of the network are set as random values within (0, 0.33). Node weights are divided into two levels with the weight of the high-level node set as 3 and that of the low-level as 1. The high/low weight represents the rank of linguistic and cultural influence of the settlements respectively. We initially set the four nodes numbered (4, 4), (4, 16), (16, 4), and (16, 16) as the high-level weighted nodes. The value of $\psi$ is set to 4, and the duration of language evolution is set to 1000 units of time. For convenience, in what follows the above settings are collectively referred to as standard initial settings.

At the initial moment, the distribution of language evolution states of each settlement in the state space is shown in Fig. 3. According to the figure, the distribution of the language state vectors of the 400 settlements in the two-dimension state space is approximately uniform.

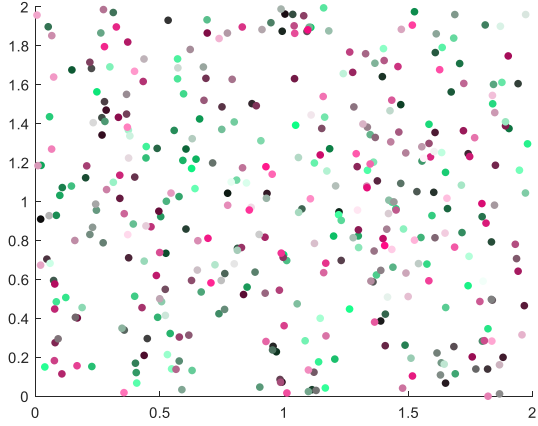

FIGURE 3. Distribution of state points within the state space at the initial time. For $(i,j)$ in the grid, the color of its state point is fixed while the position will move with time.

In the figure above, each state point has its color, and the color is set according to the index in the grid. For node $(i,j)$,

$$\begin{cases} R(i,j) = i/20 \\ G(i,j) = j/20 \\ B(i,j) = (i+j)/40 \end{cases} \qquad (23)$$

In this manner, any two nodes being adjacent in grid should also be close in color, while a short distance between state points implies a small language difference.

## B. INFLUENCE OF SELF-EVOLUTION OFFSET

In this subsection, we only investigate the influence of magnitude of the offset on the result of language evolution, with the other parameters keeping standard initial settings. In this setting, each entry of the offset vector is randomly generated in (-0.02, 0.08), and the magnitude of offset is changed by enlarging and shrinking the interval span.

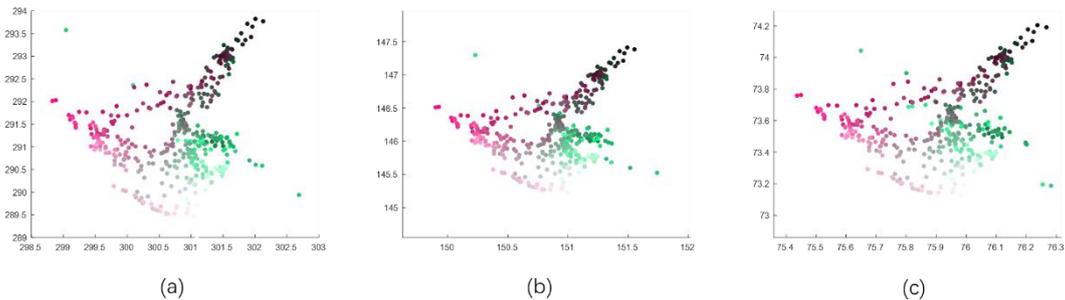

FIGURE 4. Corresponding evolution results from three different settings. As the control group, (a) is the result of language evolution under the standard initial settings. (b) is the corresponding language evolution result after the magnitude of the offset of the control group is reduced by half and (c) is that with the magnitude being reduced to a quarter of the original setting.

Then, we investigated the space occupied by the resulting image and the transfer of the location, as

shown in TABLE 1:

TABLE 1. Quantitative effect of offset shrinkage on the evolution result.

|  | (a) | (b) | (c) |
|---|---|---|---|
| Amount of space taken | 4×5 | 2×2.5 | 1×1.2 |
| Center coordinate of the image | (301,291) | (151,146) | (76,73.5) |

Note: The values in the table are calculated by approximation.

According to Fig. 4, the spatial distribution pattern of state points in the system remains similar. In order to explore its internal law, we conduct further analysis based on the data in TABLE 1. It is not difficult to find that the space occupied by state points is approximately reduced to $1/k^2$ $(k > 1)$ of that in the control group, and the center coordinate of the image also shrinks to $1/k$ of what it was when the magnitude of the offset is reduced to $1/k$ of the original value.

## C. INFLUENCE OF DISTRIBUTIONS OF NODES

In our model, high-weight nodes correspond to relatively developed settlements in reality, and it is known according to the power-law rule of distribution [33] that these nodes must only account for a small portion of settlements. This subsection mainly introduces the influence of the specific topology of the distribution of high-weight nodes on the result of language evolution. In this subsection, we mainly study two typical topologies of distribution of high-weight nodes, which are O-type and X-type respectively.

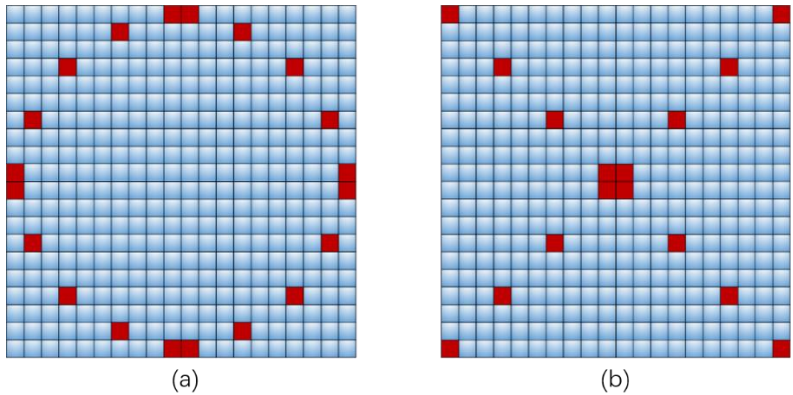

FIGURE 5. Distribution diagram of nodes with high weight. (a) is an O-type distribution. This distribution topology of nodes with high weight in the plane looks like an 'O', which is an edge-surround type distribution. (b) is an X-type distribution. This topology is similar to an 'X', which radiates from the center to the periphery.

For the two completely different distributions above, after 10000 iterations, the language evolution results are shown in Fig. 6:

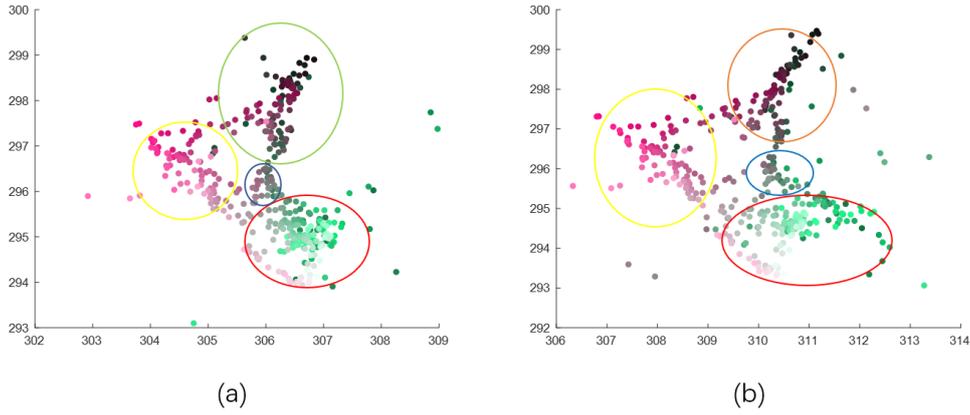

FIGURE 6. Schematic representation of language evolution results in state space. (a) and (b) are the experimental results under O-type and X-type distribution respectively.

Different from the initial condition when the state points are uniformly distributed in a certain region of the state space, the state points of the settlements form a special pattern after 10000 iterations. From Fig. 6, one can observe that two corresponding clusters may have distinct differences in shape, e.g., the two clusters composed of green dots. Nevertheless, the basic clustering configuration is unchanged, as explicitly visualized by the four clusters along with their inter-transits simultaneously appearing in both images.

## D. INCREASE IN DIMENSIONS

In the standard setting, both the state vector and the self-evolution offset are two-dimensional vectors. Here, we raise dimensions to three, leaving the rest of the settings unchanged. Compared with the two-dimensional situation, the convergence of state points according to color seems to be clearer in the three-dimensional state space.

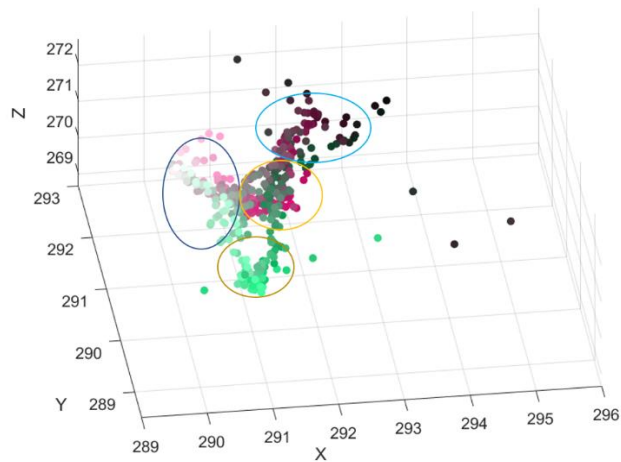

FIGURE 7. Schematic diagram of the evolution result under the condition that dimensions of state vector and self-evolving offset are increased to three dimensions.

As indicated by the colored circle box in Fig. 7, after 10,000 iterations, the 400 state points converge into four clusters according to color. It can be seen vividly that the state vectors of adjacent nodes in the topological network tend to be consistent in language more easily.

**IV. CONCLUSION**

In this paper, a multi-agent model is designed for the sake of studying the general law of language evolution with the network topology being lattice. The motion of each node is jointly driven by a fixed evolution direction itself and the influence from its neighbors in the lattice. Through a series of computational experiments, an important law of the model is revealed: after long-term language evolution of the settlements, the distribution of their state points is no more uniformly random in the state space as the initial moment, but presents a clustered formation according to their colors. Most of the state points with similar colors are clustered together, while the clusters with different colors are separated. In the model, the color of each state point depends on the index of its corresponding site in the lattice, with similarity of colors reflecting geographical proximity. This indicates that, as a result of evolution, it is universal to emerge a small number of but large-scale dialect regions within a language family. Such a rule can also be extended to other types of systems. The states in the continuous state space with initial uniformly random distribution can spontaneously organize into countable discrete clusters, which actually reflects the universality of self-organization phenomenon from disorder to order in dynamic complex network systems.

In this model, the states of languages move unperturbedly in continuous state space. This motion in reality must be stable, without the trajectories deviating far from each other. However, there is possibility for the model to be unstable, which should be avoided practically. For this end, we proved the condition of relative stability of state trajectories of adjacent nodes theoretically, to ensure that the evolution process is free from infinite divergence as a whole. This condition is also applied in the experiments.

In addition to the important discovery about clustering, our experiments also preliminarily explore the influence of the main components of the model on the evolution results, involving the self-evolution offset, the distribution of high-weight nodes in the network and the increase of dimensions of the state space. Through the analysis of the experimental results, we have a further understanding of the model. For instance, the size of the space occupied by the state point changes exponentially with the magnitude of the self-evolution offset changing.

Via a series of experiments, we have probed and discovered certain general laws of language evolution. However, in the course of research, we found that there are still some deeper contents worth studying, such as causal effect among the model components which can help to grasp the direction of evolution. Moreover, the phenomena and mechanisms of hierarchical clustering can be further studied and compared with other similar systems like the evolution of biological species. It can also be combined with specific language phenomena to refine the dynamic mechanism of node evolution and interaction between nodes, and to redefine the states into multimodal vectors. The model parameters could also be calibrated with empirical evidence if available.

## ACKNOWLEDGMENTS

This work is supported by National Natural Science Foundation (NNSF) of China (Grant 61867005), and by BUPT innovation and entrepreneurship support program (Grants 2022-YC-S006 and 2022-YC-T004).

## COMPETING INTERESTS

The authors declare that they have no competing interests regarding the publication of this paper.

## DATA AVAILABILITY

The data in this paper is generated via simulations.